\documentclass[doublespacing]{elsart}
\usepackage{amsmath,amssymb,amsfonts,graphicx,float,multicol,fleqn}
\usepackage{umoline}
\usepackage{multirow}
\newtheorem{defi}{Definition}
\newtheorem{theorem}{Theorem}
\begin{document}
\begin{frontmatter}

\title{Security analysis of two lightweight certificateless signature schemes}
\author[1]{Nasrollah Pakniat\corauthref{C.A.}}
\ead{pakniat@irandoc.ac.ir}
\corauth[C.A.]{Corresponding author. Tel:+98 2166951430; fax: +98 2166462254}
\address[1]{Information Science Research Center, Iranian Research Institute for Information Science and Technology (IRANDOC), Tehran, Iran.}

\begin{abstract}
Certificateless cryptography can be considered as an intermediate solution to overcome the issues in traditional public key infrastructure (PKI) and identity-based public key cryptography (ID-PKC). There exist a vast number of certificateless signature (CLS) schemes in the literature; however, most of them are not efficient enough to be utilized in limited resources environments such as Internet of things (IoT) or Healthcare Wireless Sensor Networks (HWSN). Recently, two lightweight CLS schemes have been proposed by Karati et al. and Kumar et al. to be employed in IoT and HWSNs, respectively. While both schemes are claimed to be existentially unforgeable, in this paper, we show that both these signatures can easily be forged. More specifically, it is shown that 1)
in Karati et al.'s scheme, a type 1 adversary, considered in certificateless cryptography, can generate a valid partial private key corresponding to any user of its choice and as a consequence, it can forge any users' signature on any message of its choice, and 2) in Kumar et al.'s scheme, both types of adversaries which are considered in certificateless cryptography are able to forge any signer's signature on an arbitrary message.

\end{abstract}


%
    \end{frontmatter}

\section{Introduction}\label{intro}
Certificateless cryptography, introduced in 2003 by Al-riyami and Paterson \cite{alriyami}, can be considered as an intermediate solution to overcome the issues in traditional public key infrastructure (PKI) and identity-based public key cryptography (ID-PKC) \cite{shamir}. Whereas a trusted authority is needed in traditional PKI to bind the identity of an entity to his public key, ID-PKC requires a trusted private key generator to generate the private keys of users based on their identities. Therefore, the certificate management problem in the public-key setting is actually replaced by the key escrow problem. In certificateless cryptography, the users' private keys are still generated with the help of a third party, called the key generation center (KGC). However, the KGC doesn't have access to the final private keys generated by the users themselves (based on some private information obtained from the KGC and some secret values chosen by the users). The public key of a user is computed from the KGC's public parameters and some information, private to the user, and is published by the user himself.
\par
Regarding the security of a certificateless cryptographic scheme, two types of adversaries are considered in the literature: a Type 1 adversary $A_1$ who simulates malicious ordinary users and a Type 2 Adversary $A_2$ who simulates a malicious KGC in a certificateless cryptographic scheme. To perform these simulations, $A_1$ is allowed to replace the public key of entities with other values of its choice and $A_2$ is allowed to get access to the master secret key.

\par
The first certificateless signature (CLS) scheme was proposed in \cite{alriyami} by Al-Riyami and Paterson. After this seminal work, a vast number of certificateless signature schemes were proposed such as ordinary CLS schemes \cite{alriyami,karati,kumar,CLS1,CLS2,CLS3,CLS4,CLS5,8546984}, certificateless proxy signature schemes \cite{CLPS1,CLPS2,CLPS3,CLPS4}, certificateless aggregate signature schemes \cite{kumar,CLAS1,CLAS2,CLAS3,CLAS4,CLAS5,clashm}, certificateless signature schemes with designated tester \cite{CLSD1,CLSD2}, certificateless threshold signature schemes \cite{CLTS1,CLTS2,CLTS3}, certificateless ring signature schemes \cite{CLRS1,CLRS2}, and etc. However, due to their heavy computational costs, most of these schemes can not be applied in limited resources environments such as Internet of Things (IoT) and Healthcare Wireless Sensor Networks (HWSN).
As a consequence, new efforts have been put forth to construct lightweight cryptographic schemes in certificateless setting in order to be applicable in limited resources environments. In this regard, recently, two lightweight certificateless signature schemes have been proposed by Karati et al. \cite{karati} and Kumar et al. \cite{kumar}. The authors of both papers claimed that their proposed CLS schemes are existentially unforgeable. However, in this paper, we disprove their claims and show that the CLS schemes of \cite{karati} and \cite{kumar} are both insecure. This is done by showing that:
\begin{itemize}
\item In \textbf{Karati et al.'s CLS scheme}, a type 1 adversary of certificateless cryptography is able to generate a valid partial private key corresponding to any identity of its choice and then uses this generated partial private key to forge the signature of the corresponding user on any message of its choice.
\item In \textbf{Kumar et al.'s CLS scheme}, both types of adversaries, considered in certificateless cryptography, are able to violate the unforgeability of the scheme. More precisely, 1) a type 1 adversary is able to forge any signer's signature on any message in this scheme as soon as it gets access to a pair of message and its corresponding signature of that signer, and 2) a type 2 adversary is able to forge each user's signature on any message in this scheme (without even requiring to see a signature of that signer).
\end{itemize}
\par
The rest of this paper is organized as follows.
In Section \ref{framework}, we provide the framework and the security definition of CLS schemes. In Section \ref{scheme1}, after reviewing the CLS scheme of \cite{karati}, we provide the proof of its insecurity. Then, the CLS scheme of Kumar et al. \cite{kumar} and analysis of its security are reviewed in Section \ref{scheme2}. Finally, the conclusions are provided in Section \ref{concl}.
\section{Certificateless signature schemes}\label{framework}
In this section, we provide the framework and the security definition of Certificateless signature schemes.
\subsection{The framework}
There exist three entities in a CLS scheme: a key generation center (KGC) which helps users to generate their private keys, a signer, and a verifier. A CLS scheme consists of six algorithms: Setup, Set-Partial-Private-Key, Set-Secret-Value, Set-Public-Key, CLS-Sign and CLS-Verify. The details of these algorithms are described in the following:
\par
\textbf{Setup}: Performed by $KGC$.
\begin{itemize}
    \item Input: The security parameter $k$.
    \item Process:
        \begin{itemize}
            \item Generates the master secret key $MSK$, and the public parameters $params$.
        \end{itemize}
    \item Output: The master secret key $MSK$ which will be secured by KGC and the public parameters $params$ which are published.
\end{itemize}

\textbf{Set-Partial-Private-Key}: Performed by $KGC$.
\begin{itemize}
  \item Input: $params$, $MSK$ and a user's identity $ID_S$.
  \item Process:
  \begin{itemize}
  \item Computes a partial private key $D_S$ corresponding to this user.
  \end{itemize}
  \item Output: Partial private key $D_S$ which will be sent securely to the user with identity $ID_S$.
\end{itemize}

\textbf{Set-Private-Key}: Performed by a user $S$.
  \begin{itemize}
    \item Input: $params$ and $S$'s partial private key $D_S$.
    \item Process:
    \begin{itemize}
    \item Generates a secret value $x_S$ and computes the private key $SK_S$ by using it and $D_S$.
    \end{itemize}
    \item Output: $SK_{S}$ which will be secured by the user $S$.
  \end{itemize}

\textbf{Set-Public-Key}: Performed by a user $S$.
  \begin{itemize}
    \item Input: $params$ and $S$'s private key $SK_S$.
    \item Process:
    \begin{itemize}
    \item Computes the public key $PK_S$.
    \end{itemize}
    \item Output: $PK_S$ which will be published.
  \end{itemize}

\textbf{CLS-Sign}: Performed by the user $S$.
\begin{itemize}
    \item Input: $params$, the user's identity $ID_S$ and his private key $SK_S$, and a message $m$.
    \item Process:
    \begin{itemize}
        \item Generates a signature $\sigma$ on the message $m$.
    \end{itemize}
    \item Output: $\sigma$ as the signature on $m$.
\end{itemize}

\textbf{CLS-Verify}: Performed by the verifier.
\begin{itemize}
    \item Input: $params$, signer's identity $ID_S$ and his public key $PK_{S}$, message $m$ and a signature $\sigma$.
    \item Process:
    \begin{itemize}
        \item Checks the validity of $\sigma$.
    \end{itemize}
      \item Output: VALID if $\sigma$ is a valid signature on $m$ and INVALID otherwise.
    \end{itemize}

\subsection{Security model}
To call a CLS scheme secure, it should provide existentially unforgeability against adaptive chosen-message and -identity attacks in the adversarial model of certificateless cryptography which consists of the following two types of adversaries:
\begin{itemize}
\item A type-1 adversary ($A_1$), that has not access to the master secret key but can replace any signer's public key with any value of its choice.
\item A type-2 Adversary ($A_2$), that has access to the master secret key but cannot replace public keys.
\end{itemize}
The security of a CLS scheme is modeled through the following two games played between a challenger $C$ and adversaries $A_1$ or $A_2$.
\par
\textbf{Game 1}: This game, played between $C$ and $A_1$, consists of the following phases:
\begin{itemize}
\item Setup: In this phase, $C$ generates the master secret key $MSK$ and the public parameters $params$. It keeps $MSK$ secure and sends $params$ to $A_1$.
\item Queries: In this phase, $A_1$ can perform a polynomially bounded number of the following queries and $C$'s answers to these queries are as follows:
\begin{itemize}
\item Request-Partial-Private-Key ($ID_S$): inputting $ID_S$ to this query, $A_1$ will get $S$'s partial private key $D_S$ as the output.
\item Request-Secret-Value ($ID_S$): inputting $ID_S$ to this query, $A_1$ will get $S$'s secret value $x_S$ as the output.
\item Request-Public-Key ($ID_S$): inputting $ID_S$ to this query, $A_1$ will get $S$'s public key $PK_S$ as the output.
\item Replace-Public-Key ($ID_S,PK^\prime_S$): inputting $ID_S$ and $PK^\prime_S$ to this query, $PK^\prime_S$ will be set as the public key corresponding to the user $S$.
\item CL-Sign ($ID_S,m$): inputting $ID_S$ and $m$ to this query, $A_1$ will get $\sigma$ as the output which is a valid signature of $S$ on $m$.
\end{itemize}
\item Output: Finally, when $A_1$ decides to end the queries phase, it outputs a signature $\sigma$ on a message $m$ on behalf of a targeted user with identity $ID$. It wins the game if the following conditions are fulfilled:
\begin{itemize}
\item The algorithm CLS-Verify outputs VALID on inputs $params$, $m$, $\sigma$, $ID$, and $PK$ where, $PK$ is the public key corresponding to the user with identity $ID$.
\item The queries Request-Partial-Private-Key($ID$) and CL-Sign($ID,m$) weren't queried in the queries phase.
\end{itemize}
\end{itemize}
\begin{defi} A CLS scheme is Type-1 secure against the adaptively chosen-message and -identity attack if the advantage of any polynomially bounded adversary $A_1$  in winning Game 1 be negligible.
\end{defi}
\textbf{Game 2}: This game, played between $C$ and $A_2$, consists of the following phases:
\begin{itemize}
\item Setup: In this phase, $C$ generates the master secret key $MSK$ and the public parameters $params$ and sends them to $A_2$.
\item Queries: In this phase, $A_2$ can perform a polynomially bounded number of queries as in Game 1 and $C$ answers them in the same way. The only constraint here is that $A_2$ is not allowed to replace any public keys. Note that $A_2$ knows $MSK$ and can compute the partial private key of any identity by itself.
\item Output: Finally, when $A_2$ decides to end the queries phase, it outputs a signature $\sigma$ on a message $m$ on behalf of a targeted user with identity $ID$. It wins the game if the following conditions are fulfilled:
\begin{itemize}
\item The algorithm CLS-Verify outputs VALID on inputs $params$, $m$, $\sigma$, $ID$, and $PK$ where, $PK$ is the public key corresponding to the user with identity $ID$.
\item The queries Request-Secret-Value($ID$) and CL-Sign($ID,m$) weren't queried in the queries phase.
\end{itemize}
\end{itemize}
\begin{defi}
A CLS scheme is Type-2 secure against the adaptively chosen-message and -identity attack if the advantage of any polynomially bounded adversary $A_2$ in winning Game 2 be negligible.
\end{defi}
\section{Karati et al.'s CLS scheme}\label{scheme1}
In this section, we first review Karati et al.'s CLS scheme and then prove that it is completely insecure.
\subsection{Review of the scheme}\label{review1}
The CLS scheme of Karati et al. \cite{karati} consists of the following algorithms:
\par
\textbf{Setup}: Performed by $KGC$.
\begin{itemize}
    \item Input: The security parameter $k$.
    \item Process:
        \begin{itemize}
            \item Generates two groups $G_1$ and $G_2$ with the same prime order $p$ and an efficient bilinear pairing $e:G_1\times G_1\rightarrow G_2$.
            \item Chooses a generator $g_1 \in G_1$.
            \item Chooses a cryptographic hash function $H: \{0,~1\}^\ast\rightarrow Z^\ast_p$.
            \item Chooses a random $y\in Z^\ast_p$ as his master secret key.
            \item Computes $g_2=e(g_1,g_1)^y$ and $Y_{KGC}=g_1^y$.
        \end{itemize}
    \item Output: The master secret key $y$ which will be secured by KGC and the public parameters $params=(G_1,~ G_2,~p,~ e,~ g_1,~g_2,~ Y_{KGC},~ H)$ which will be published.
\end{itemize}
\textbf{Set-Partial-Private-Key}: Performed by $KGC$.
\begin{itemize}
  \item Input: $params$, master secret key $y$ and a user's identity $ID_i \in \{0,~1\}^\ast$.
  \item Process:
  \begin{itemize}
  \item Computes $h_i = H(ID_i)$.
  \item Chooses $r_i\in Z^\ast_p$ randomly and computes $R_i=g_1^{r_i}$ and $y_i=\left(g_1\right)^{\frac{y\cdot h_i}{h_i+r_i+y}}$.
  \end{itemize}
  \item Output: Partial private key $D_i=(y_i,R_i)$ which will be sent securely to the user with identity $ID_i$. After receiving $D_i$ from $KGC$, the user considers $D_i$ genuine if:
      \begin{eqnarray}
      e(g_1,Y_{KGC})^{h_i}=e(y_i,(g_1^{h_i}\cdot R_i\cdot Y_{KGC})).
      \end{eqnarray}
\end{itemize}
\textbf{Set-Private-Key}: Performed by a user $i$.
  \begin{itemize}
    \item Input: $params$ and $i$'s partial private key $D_i=(y_i,R_i)$.
    \item Process:
    \begin{itemize}
    \item Chooses $x_i,c_i\in Z^\ast_p$ randomly and sets $SK_i=(c_i,x_i,R_i)$.
    \end{itemize}
    \item Output: $SK_{i}$ which will be secured by the user $i$.
  \end{itemize}

\textbf{Set-Public-Key}: Performed by a user $i$.
  \begin{itemize}
    \item Input: $params$, $i$'s partial private key $D_i=(y_i,R_i)$ and his private key $SK_i=(c_i,x_i,R_i)$.
    \item Process:
    \begin{itemize}
    \item Computes $Y_i = \left(Y_{i1}=(y_i)^{\frac{1}{x_i}},Y_{i2}=g_2^{c_i}\right)$ as the user's public key.
    \end{itemize}
    \item Output: $Y_i$ which will be published.
  \end{itemize}
\textbf{CLS-Sign}: Performed by a user $S$.
\begin{itemize}
    \item Input: $params$, the user's identity $ID_S$ and his private key $SK_S=(c_S,x_S,R_S)$ and a message $m$.
    \item Process:
    \begin{itemize}
        \item Computes $h_S=H(ID_S)$.
        \item Chooses a random value $t\in Z^\ast_p$ and computes
        \begin{eqnarray}
        \sigma_1&=&g_2^t,\\
        \sigma_2&=&\left(g_1^{h_S}\cdot R_S \cdot Y_{KGC}\right)^{\left(\frac{c_S}{m}-t\right)x_S}.
        \end{eqnarray}
    \end{itemize}
    \item Output: $\sigma=(\sigma_1,\sigma_2)$ as the signature on $m$.
\end{itemize}
\textbf{CLS-Verify}: Performed by the verifier.
\begin{itemize}
    \item Input: $params$, $S$'s identity $ID_S$ and his public key $Y_{S}=(Y_{S1},Y_{S2})$, message $m$ and a signature $\sigma = (\sigma_1,\sigma_2)$.
    \item Process:
    \begin{itemize}
        \item Computes $h_{S}= H(ID_S)$.
        \item Checks whether $\left(\frac{Y_{S2}^\frac{1}{m}}{\sigma_1}\right)^{h_S}=^?e\left(Y_{S1},\sigma_2\right)$ .
    \end{itemize}
      \item Output: VALID if the above equation holds and INVALID otherwise.
    \end{itemize}

\subsection{Cryptanalysis of the scheme}\label{cryptanalysis1}
The authors of \cite{karati} claimed that their proposed scheme is a secure certificateless signature scheme. However, in this section, we disprove their claim. More specifically, we show that by accessing to a valid partial private key corresponding to any user, a valid partial private key corresponding to any other user can be generated. Thereupon, each user of this scheme can forge the signature of other users on any arbitrary message of his choice.
This is formally stated and proved in the following theorem.
\begin{theorem}
Let $S$ with identity $ID_S$ be an arbitrary user of Karati et al.'s scheme. Suppose that $A_1$ has access to a valid partial private key corresponding to $S$. Then, $A_1$ is able to generate a valid partial private key corresponding to any other user $S^\prime$ with arbitrary identity $ID_{S^\prime}$ and as a consequence, he is able to forge $S^\prime$'s signature on any message of his choice.
\end{theorem}
\textbf{Proof.}
According to Set-Partial-Private-Key algorithm of Karati et al.'s CLS scheme, the partial private key corresponding to $S$ with identity $ID_S$ is a pair $(y_S,R_S)$ where, $R_S=g_1^{r_S}$ and $y_i=\left(g_1\right)^{\frac{y\cdot h_S}{h_S+r_S+y}}$ in which $r_S\in Z^\ast_p$ is an unknown randomly chosen value, $y$ is the master secret key and $h_S=H(ID_S)$.
In the following, we show how $A_1$ is able to use $S$'s partial private key to generate a valid partial private key corresponding to any other user $S^\prime$ with arbitrary identity $ID_{S^\prime}$. To this end, $A_1$:
\begin{enumerate}
\item Computes $h_{S^\prime}=H(ID_{S^\prime})\in Z^\ast_p$.
\item Computes $\alpha=\frac{h_{S^\prime}}{h_S}\in Z^\ast_p$. Note that the output of $H(\cdot)$ is a member of $Z^\ast_p$ and therefore, $h^{-1}_S$ exists in $Z^\ast_p$.
\item Computes $y_{S^\prime}=y_S^\alpha$ and $R_{S^\prime}=\frac{R_S}{g_1^{(\alpha-1)\cdot h_S}}$.
\item Sets $(y_{S^\prime},R_{S^\prime})$ as the partial private key corresponding to the user $S^\prime$ with identity $ID_{S^\prime}$.
\end{enumerate}
Using the following relation, it can easily be verified that $(y_{S^\prime},R_{S^\prime})$ is a valid partial private key corresponding to the user $S^\prime$ with identity $ID_{S^\prime}$:
\begin{eqnarray}
      &&e(y_{S^\prime},(g_1^{h_{S^\prime}}\cdot R_{S^\prime}\cdot Y_{KGC}))\\
      &=&e(y_S^\alpha,(g_1^{\alpha\cdot h_S}\cdot \frac{R_S}{g_1^{(\alpha-1)\cdot h_S}} \cdot Y_{KGC}))\\
      &=& e(y_S^\alpha,(g_1^{\alpha\cdot h_S-(\alpha-1)h_S}\cdot R_S \cdot Y_{KGC}))\\
      &=& e(y_S^\alpha,( g_1^{h_S}\cdot R_S \cdot Y_{KGC}))\\
      &=& e(y_S,( g_1^{h_S}\cdot R_S \cdot Y_{KGC}))^\alpha\\
      &=& e(g_1, Y_{KGC})^{h_{S}\cdot\alpha}\label{nexttothelast}\\
      &=& e(g_1, Y_{KGC})^{h_{S^\prime}},
\end{eqnarray}
where, equality (\ref{nexttothelast}) is obtained from the fact that $(y_S,R_S)$ is a valid partial private key generated by the $KGC$ and therefore,
\begin{eqnarray}
e(g_1,Y_{KGC})&=&e(y_S,(g_1^{h_S}\cdot R_S\cdot Y_{KGC})).
\end{eqnarray}
After computing $S^\prime$'s partial private key, $A_1$ can perform Set-Private-Key and Set-Public-key (as explained in Katari et al.'s CLS scheme) instead of $S^\prime$ to compute a valid pair of private and public keys corresponding to $S^\prime$. Now, using the private key of $S^\prime$, $A_1$ can forge $S^\prime$'s signature through CLS-Sign algorithm on any message of its choice.
\hfill $\Box$
\section{Kumar et al.'s CLS scheme}\label{scheme2}
In this section, we first review Kumar et al.'s CLS scheme and then prove that their scheme is forgeable.
\subsection{Review of the scheme}\label{review2}
The CLS scheme of Kumar et al. \cite{kumar} consists of the following algorithms:
\par
\textbf{Setup}: Performed by $KGC$.
\begin{itemize}
    \item Input: The security parameter $k$.
    \item Process:
        \begin{itemize}
            \item Chooses two groups $G_1$ and $G_2$ with the same prime order $q$ and a generator $P$ in $G_1$.
            \item Chooses a bilinear map $e:G_1\times G_1\rightarrow G_2$.
            \item Chooses a random $\alpha \in Z^\ast_q$ as the master secret key and sets $P_{Pub} = \alpha\cdot P$.
            \item Chooses cryptographic hash functions $H_1,H_2: \{0,~1\}^\ast\rightarrow G_1$ and $H_3:\{0,~ 1\}^\ast\rightarrow Z^\ast_q$.
        \end{itemize}
    \item Output: The master secret key $\alpha$ which will be secured by $KGC$ and the system parameters $params=(q,~ G_1,~ G_2,~ e,~ P,~ P_{Pub},~ H_1,~ H_2,~H_3)$ which will be published.
\end{itemize}
\textbf{Set-Partial-Private-Key}: Performed by $KGC$.
\begin{itemize}
  \item Input: $params$, master secret key $\alpha$ and a user's identity $ID_i \in \{0,~1\}^\ast$.
  \item Process:
  \begin{itemize}
  \item Computes $Q_{ID_i} = H_1(ID_i)$.
  \item Computes $D_{i} = \alpha\cdot Q_{ID_i}$.
  \end{itemize}
  \item Output: Partial private key $D_{i}$ which will be sent securely to the user with identity $ID_i$.
\end{itemize}
\textbf{Set-Private-Key}: Performed by a user $i$.
  \begin{itemize}
    \item Input: $params$ and $i$'s identity $ID_i$.
    \item Process:
    \begin{itemize}
    \item Selects a random value $x_{i} \in Z^\ast_q$ as the $i$'s secret key.
    \item Sets $SK_i=(x_i,D_i)$.
    \end{itemize}
    \item Output: $SK_i$ which will be secured by the user $i$.
  \end{itemize}

\textbf{Set-Public-Key}: Performed by a user $i$.
  \begin{itemize}
    \item Input: $params$ and $i$'s private key $SK_i=(x_i,D_i)$.
    \item Process:
    \begin{itemize}
    \item Computes $Y_{i} = x_i\cdot P$ as $i$'s public key.
    \end{itemize}
    \item Output: $Y_i$ which will be published.
  \end{itemize}

\textbf{CLS-Sign}: Performed by a user $S$.
\begin{itemize}
    \item Input: $params$, the signer's identity $ID_S$, his public key $Y_{S}$, his private key $SK_S=(x_S,D_S)$, some state information $\Delta$ and a message $m$.
    \item Process:
    \begin{itemize}
        \item Chooses a random value $r\in Z^\ast_q$ and computes $R=r\cdot P\in G_1$.
        \item Computes $W=H_2(\Delta)$ and $h=H_3(m,~ID_S,~Y_S,~R)$.
        \item Computes $V=D_S+r\cdot W+h\cdot x_S \cdot P_{Pub}$.
    \end{itemize}
    \item Output: $\sigma=(R,~V)$ as the signature on $m$ under the state information $\Delta$.
\end{itemize}
\textbf{CLS-Verify}: Performed by the verifier.
\begin{itemize}
    \item Input: $params$, signer's identity $ID_S$ and his public key $Y_S$, message $m$, some state information $\Delta$ and a signature $\sigma = (R,~V)$.
    \item Process:
    \begin{itemize}
        \item Computes $Q_{ID_S}= H_1(ID_S)$, $W=H_2(\Delta)$ and $h=H_3(m,~ID_S,~Y_S,~R)$.
        \item Verifies $e(V,~P)=^?e(Q_{ID_S}+h\cdot Y_S,P_{Pub})e(R,W)$.
    \end{itemize}
      \item Output: VALID if the above equation holds and INVALID otherwise.
    \end{itemize}

\subsection{Cryptanalysis of the scheme}\label{cryptanalysis2}
Kumar et al. claimed that their scheme is existentially unforgeable against adaptive chosen message attacks. However, in this section, we disprove their claim.
We prove the insecurity of Kumar et al.'s CLS scheme by the following theorems:


\begin{theorem}
Let $S$ be a signer with identity $ID_S$ who uses Kumar et al.'s CLS scheme. Suppose that a type 1 adversary $A_1$ has access to a tuple $(m,~\sigma=(R,V)~,\Delta)$, where $\sigma$ is $S$'s signature on message $m$ under the state information $\Delta$. Then, $A_1$ is able to forge $S$'s signature on any new message $m^\prime$ under the same state information $\Delta$.
\end{theorem}
\textbf{Proof.}
According to Kumar et al.'s CLS-Sign algorithm, the signature $\sigma$ is as follows:
    \begin{eqnarray}
    R=r\cdot P,~~~~~V=D_S+r\cdot H_2(\Delta)+x_S\cdot h\cdot P_{Pub},
    \end{eqnarray}
    where $h=H_3(m,~ID_S,~Y_S,~R)$ and $r\in Z^\ast_q$ is a random value that is unknown to $A_1$. Now, in order to forge $S$'s signature on a new massage $m^\prime$, $A_1$:

\begin{enumerate}
\item Issues a Request-Secret-Value query on the input of $ID_S$ and obtains $x_S$ as the result.
\item Computes $D_{S,\Delta}= V- x_S\cdot h\cdot P_{Pub}=D_S+ r\cdot H_2(\Delta)$.
\item Uses $D_{S,\Delta}$, $x_S$ and $R$ to forge $S$'s signature on $m^\prime$ as follows:
    \begin{enumerate}
        \item[1.] Computes $h^\prime=H_2(m^\prime,~ID_S,~Y_S,~R)$ and $V^\prime=D_{S,\Delta}+h^\prime\cdot x_S \cdot P_{Pub}$.
        \item[2.] Outputs $\sigma^\prime=(R,~V^\prime)$ as $S$'s signature on message $m^\prime$.
    \end{enumerate}
It can be easily verified that the forged signature $\sigma^\prime$ is valid.
\end{enumerate}
\hfill $\Box$

\begin{theorem}
Let $S$ be a signer with identity $ID_S$ who uses Kumar et al.'s CLS scheme. Then, a type 2 adversary $A_2$ is able to forge $S$'s signature on any message $m$ of its choice under any arbitrary state information $\Delta$.
\end{theorem}
\textbf{Proof.}
To forge $S$'s signature on any arbitrary message $m$, $A_2$:
    \begin{enumerate}
        \item[1.] Chooses a random value $r\in Z^\ast_q$ and computes $R=r\cdot P$.
        \item[2.] Computes $h=H_3(m,~ID_S,~Y_S,~R)$ and $V=D_S+r H_2(\Delta)+h\cdot \alpha \cdot Y_S$.
        \item[3.] Outputs $\sigma=(R,~V)$ as $S$'s signature on message $m^\prime$.
    \end{enumerate}
Note that $A_2$ acts as the malicious key generation center and has access to partial private keys. It can be easily verified that the forged signature $\sigma$ is valid.
    \hfill $\Box$

\section{Conclusion}\label{concl}
In this paper, the security of two recently proposed lightweight certificateless signature schemes is considered. We prove that in one of them, a type 1 adversary of certificateless cryptography can forge the signature of any user on any arbitrary message of his choice and in the other one, both considered types of adversaries in certificateless cryptography can forge valid signatures on behalf of any user on any message of their choices.

\end{document}